\documentclass[namedreferences]{solarphysics}
\usepackage[pdfborder={0 0 0 },urlcolor=blue,breaklinks]{hyperref}
\usepackage[optionalrh,natbib]{spr-sola-addons} 
\usepackage{graphicx}        
\usepackage{color}           
\usepackage{breakurl}             

\newcommand{\etal}{{\it et al.}}

\newcommand{\annG}{   {\it Ann. Geophys.}}

\begin{document}

\begin{article}

\begin{opening}

\title{Tracking Back the Solar Wind to its Photospheric Footpoints from \emph{Wind} Observations ---
A Statistical Study}

\author{Chong Huang$^{1}$\sep
        Yihua Yan$^{1}$\sep
        Gang Li$^{1,2}$\sep
        Yuanyong Deng$^{1}$\sep
        Baolin Tan$^{1}$}
\runningauthor{C. Huang \etal}
\runningtitle{Tracking Back Solar Wind to Its Photospheric Footpoints}

   \institute{$^{1}$ Key Laboratory of Solar Activity, National Astronomical Observatories of Chinese Academy of Sciences, Beijing, China, 100012\\ email: \url{chuang@nao.cas.cn} \\
              $^{2}$ Department of Physics and CSPAR, University of Alabama in Huntsville, AL, 35899, USA
             }

\begin{abstract}

It is of great importance to track the solar wind back to its photospheric
source region and identify the related current sheets; this will
provide key information for investigating the origin and
predictions of the solar wind. We report a statistical study relating the photospheric footpoint motion and
{\it in-situ} observation of current sheets in the solar wind.
We used the potential force-free source--surface (PFSS) model and
the daily synoptic charts to trace the solar wind back from 1 AU, as observed by
the {\it Wind} spacecraft, to the solar surface. As the footpoints move along the solar surface we obtain a time series of
the jump times between different points. These jumps can
be within a cell and between adjacent cells.
We obtained the distribution of the jump times and the distribution for a subset
of the jump times in which only jumps between adjacent cells were counted.
For both cases, the distributions clearly show two populations. These distributions are compared with the distribution of {\it in-situ} current sheets reported in an earlier work of \citeauthor{mia11} (\annG\ \textbf{29}, 237, \citeyear{mia11}). Its implications on the origin of the current sheets are discussed.

\end{abstract}
\keywords{Solar wind, Magnetic fields, Current Sheets, Supergranulation}
\end{opening}

\section{Introduction}

Magnetohydrodynamic (MHD) turbulence intermittency in the solar wind
is an important topic in space plasma research
\cite{tu1995,Goldstein1995,bru05}. The intermittency emerges because the magnetic fields and
velocity fluctuations are not scale invariant as conjectured in
the hydrodynamic turbulence theory of \inlinecite{kol41}. An
important intermittent structure in the solar wind is the current sheet,
which is a two-dimensional structure where the magnetic-field
directions change significantly. Earlier work on the existence
of current sheets and their effects on solar-wind MHD turbulence
has been carried out by \inlinecite{Vel99}. These authors applied the
Haar wavelets technique to calculate the power spectra and
structure functions of the solar wind using fluid velocity and
magnetic-field data from the {\it International Sun-Earth Explorer}
(ISEE) space experiment. The temporal separations in their study range
from one minute to about one day. They suggested that the most important
intermittent structures in the solar wind are shocks and
current-sheet-like structures.

\inlinecite{bru01} studied current sheets using {\it Helios 2} data
and suggested for the first time that current sheets were
ubiquitous in the solar wind and probably are boundaries of flux tubes.
In a subsequent work, \inlinecite{bru04} furthermore obtained the
probability distribution functions (PDF) of the solar-wind
fluctuations and confirmed the results of \inlinecite{bru01}: there appeared to be two populations of current sheets in the solar
wind, one of which was flux-tube boundaries.

Alternative views about the origin of current sheets also exist.
For example, numerical MHD simulations by
\inlinecite{zhou04} and \inlinecite{chang04} showed that
current sheets can emerge from the dynamical evolution of the
nonlinear interactions of the solar-wind MHD turbulence. All of
these studies suggested that the current sheet is an intrinsic
property of the solar-wind MHD turbulence. In contrast,
\inlinecite{bor08} examined one-year magnetic-field data from the {\it Advanced Composition Explorer} (ACE) spacecraft and found that the population of the angle
between two magnetic field measurements with a separation of 64 seconds
showed a clear signature of two populations, supporting the
earlier claim of \inlinecite{bru01}. \inlinecite{bor08} suggested
that these current sheets are ``magnetic walls'' of flux tubes in
the solar wind and are relic structures that can be traced back
to the surface of the Sun. In this scenario, current sheets are
carried outward by the solar wind as passive structures. The plasma in
the solar wind is bundled in ``spaghetti-like'' flux tubes
\cite{bar66,mcc66,mar73}.

Recently, \inlinecite{tre2013b} analyzed the {\it in-situ} observations
of solar energetic particles (SEPs) and found several SEP
modulation signatures in local magnetic field and/or plasma
parameters. From studying the magnetic helicity, it is possible
to identify magnetic boundaries associated with variations of
plasma parameters, which are thought to represent the borders
between adjacent magnetic-flux tubes. The authors found that SEP
dispersionless modulations are generally associated with such
magnetic boundaries. They also analyzed the local magnetic field topology in depth by applying a Grad¨C-Shafranov
reconstruction \cite{tre2013b}, and found that flux ropes or current sheets with a more
complex field topology are generally associated with the maxima
in the SEP counts. This association shows that the SEPs propagate
within these structures and cannot escape from them because of
their much smaller gyration radii relative
to the transverse dimensions of those structures \cite{tre2013a}.

To identify current sheets in the solar wind, \citeauthor{li07} (\citeyear{li07,li08a}) developed a systematic method to obtain
the exact locations of individual current sheets.
\inlinecite{li08b} applied this method to data of the
\textit{Cluster} spacecraft in an attempt to identify the origin
of these current sheets. The \textit{Cluster}
spacecraft was chosen because its orbit traverses both the solar wind and Earth's magnetosphere. \inlinecite{li08b} found that unlike in the solar wind, there is no clear signature of current sheets in Earth's magnetosphere. This result is a natural outcome when
current sheets of solar wind are indeed relic structures
originating from the solar surface. Extending the work of
\citeauthor{li07} (\citeyear{li07,li08a}), \inlinecite{mia11} developed an automatic current-sheet identification procedure and
examined the solar-wind magnetic-field data of \textit{Ulysess}.
Their results are consistent with those of \inlinecite{bor08}.

To confirm whether current sheets are boundaries of flux tubes that originate from the solar surface, one straightforward approach
is to trace the solar wind back from {\it in-situ} observations near
Earth's orbit at 1 AU to the solar surface and examine if, for those
current sheets observed {\it in-situ}, there are corresponding footpoint
jumps between adjacent photospheric cells. To identify these
photospheric cells, \inlinecite{hua12} used a watershed algorithm
and the magnetogram observed by the {\it Solar and Heliospheric
Observatory} (SOHO)/{\it Michelson Doppler Imager} (MDI: \inlinecite{scherrer1995}). After identifying the photospheric cells, one
can then examine how the footpoints jump within and between these
cells. If the two different jump times have
different distributions, we may deduce that the current sheets
originate from two different mechanisms, and when the two different
jump times have the same distribution, the same generation
mechanism of the current sheets can be considered.

The solar wind from near Earth's orbit to the
solar surface has been back-traced before. For example, to study the
quadrupole distortions of the heliospheric current sheet and
compare the K-coronameter observations with a potential-field model,
\citeauthor{bru82} (\citeyear{bru82,bru84}) used the solar-wind tracing-back process
during the period from May 1976 to May 1977. \inlinecite{neu98}
used the data obtained by \textit{Ulysess} and \textit{Wind} in
early 1995 to trace back solar wind structures to the source surface and
then map them back to the photosphere. They found that the footpoints of
the open field-lines calculated from the model are generally
consistent with observations in the He $10\,830$ {\AA} line of
locations of coronal holes. Recently, in an attempt to make a
physical connection between the Equatorial Coronal Hole (ECH) and
the solar wind observed at about 1 AU, \inlinecite{mci10} used the
data from ACE to track the solar wind back to the solar
surface using the Potential Field Source Surface (PFSS) model
\cite{sch69,alt69}.

In mapping the solar wind back to the Sun, it was assumed that the
coronal magnetic field is quasi-stationary \cite{sch69,alt69}. This allows one to apply the PFSS model using
one synoptic chart. However, such methods will inevitably produce
some artificial bias towards the two sides of the magnetogram,
because some source regions and magnetic-field lines do not last
for one Carrington Rotation (CR). In a recent work,
\inlinecite{sun2012} used the nonlinear force-free field (NLFFF)
extrapolation to demonstrate that the change in the photospheric
and coronal field is morphologically consistent with the
``magnetic implosion'' conjecture; this change is supported by the
coronal-loop retraction observed by the {\it Solar Dynamic Observatory}
(SDO)/{\it Atmospheric Imaging Assembly} (AIA: \inlinecite{lemen2012}).

Instead of using a single synoptic chart, we used in our study the
daily synoptic chart as the boundary for the PFSS model. This
method improves the accuracy of the two sides of the synoptic
chart, and consequently the locations of the photospheric
footpoints. We applied this revised model to map the solar wind
back to the solar surface using {\it in-situ} data from {\it Wind} and ACE observations, and then calculated the jump
times between adjacent footpoints, and the jump times of
adjacent footpoints that jump across the boundaries of cells.

If the current sheets in solar wind are relic structures
that originate from the solar surface, one can expect that the waiting-time
statistics of these {\it in-situ} current sheets and that of the jump
time between adjacent cells is similar. Likewise, the
distribution between all jump times and that of footpoints
that cross the boundaries of cells will be similar. In this work
we performed such a statistical comparison. To our knowledge, this
is the first attempt of a semi-quantitative study to relate
the solar-surface observations with {\it in-situ} solar wind
observations. The article is organized as follows: Section 2
introduces the data analysis procedure and the data reduction, the
results are presented in Section 3, and the discussion and main
conclusions are presented in Section 4.

\section{Observations and Data Reduction}

\subsection{Data Selection}

To accurately track the solar wind back from near Earth's orbit (at
about 1 AU) to the solar source surface ({\it e.g.} 2.5 R$_\odot$), in the
extrapolation of the coronal magnetic field, we generally take the 2.5
R$_\odot$ for the solar source surface, where R$_\odot$ is the radius of
the Sun. Therefore, it is necessary to select quasi-stationary periods
of the solar wind to perform our analysis.

We focus on studying the minimum of the solar cycle when the
solar wind and solar magnetic field can be regarded as
quasi-stationary (\opencite{bru82}, \citeyear{bru84}; \opencite{neu98}); furthermore, transient disturbances, such as CMEs, are relatively inactive
\cite{mia11}. We selected the data during the declining phase of Solar Cycle 23 from 2004 to 2005 (CR2012\,--\,CR2037) obtained by the
ACE and {\it Wind} spacecrafts. The ACE
spacecraft is located at the Lagrangian point (The Sun--Earth
$L_1$), while the {\it Wind} spacecraft spends some of its time
within Earth's magnetosphere. For ACE, the plasma data were
obtained by the {\it Solar Wind Electron, Proton and Alpha Monitor}
(SWEPAM), and the magnetic-field data were obtained by the {\it Magnetometer}
(MAG: \inlinecite{mcc98}). For {\it Wind}, the plasma and magnetic-field data
were obtained by the {\it Solar Wind Experiment} (SWE) and the {\it Magnetic
Field Investigation} (MFI) \cite{ogi95,lep95}, respectively.

The present analysis is based on one-hour averages of
the {\it in-situ} observation data.

\subsection{Tracking the Solar Wind from {\it In-Situ} Back to the Source Surface}

We discuss the back-tracking procedure of the solar wind from 1 AU
observation to the source surface. We assumed that the source surface is
located at about 2.5 R$_\odot$ from the center of the Sun.

To obtain the accurate positions of footpoints at the source
surface, we first transformed the position of the spacecraft ACE and {\it Wind} from the Geocentric Solar Ecliptic
system (GSE) to the Carrington coordinate system \cite{hap92} and
obtained the heliographic latitude and longitude of {\it Wind}
and ACE. The latitude of the
footpoint is approximately equal to the heliographic latitude of the
spacecraft, and the longitude of the footpoint is then obtained
by taking into account the effect of solar-wind propagation.

We calculated the propagation time in the Heliocentric Earth
Ecliptic (HEE) coordinate system where the $x$-axis is along the
Sun--Earth line and the $z$-axis points to the ecliptic north pole. $P_x$ denotes as the distance between the spacecraft and the center
of the Sun, $V_x$ the radial velocity of the solar wind
observed at the spacecraft. We furthermore assumed $V_x$ to be a constant
during the propagation. The propagation time of the solar wind
from the spacecraft to the source surface is then

\begin{equation}
\Delta t=\frac{P_x-2.5R_s}{V_x},
\end{equation}
and the offset to the longitude produced from the solar wind propagation is
\begin{equation}
D=\frac{360^\circ}{27.2753\times86400}\Delta t.
\end{equation}

Adding this offset, we obtain the actual positions of footpoints
that track back from {\it in-situ} to the solar source surface.

\begin{figure}[ht]
\centering
\includegraphics[width=0.49\textwidth,clip=]{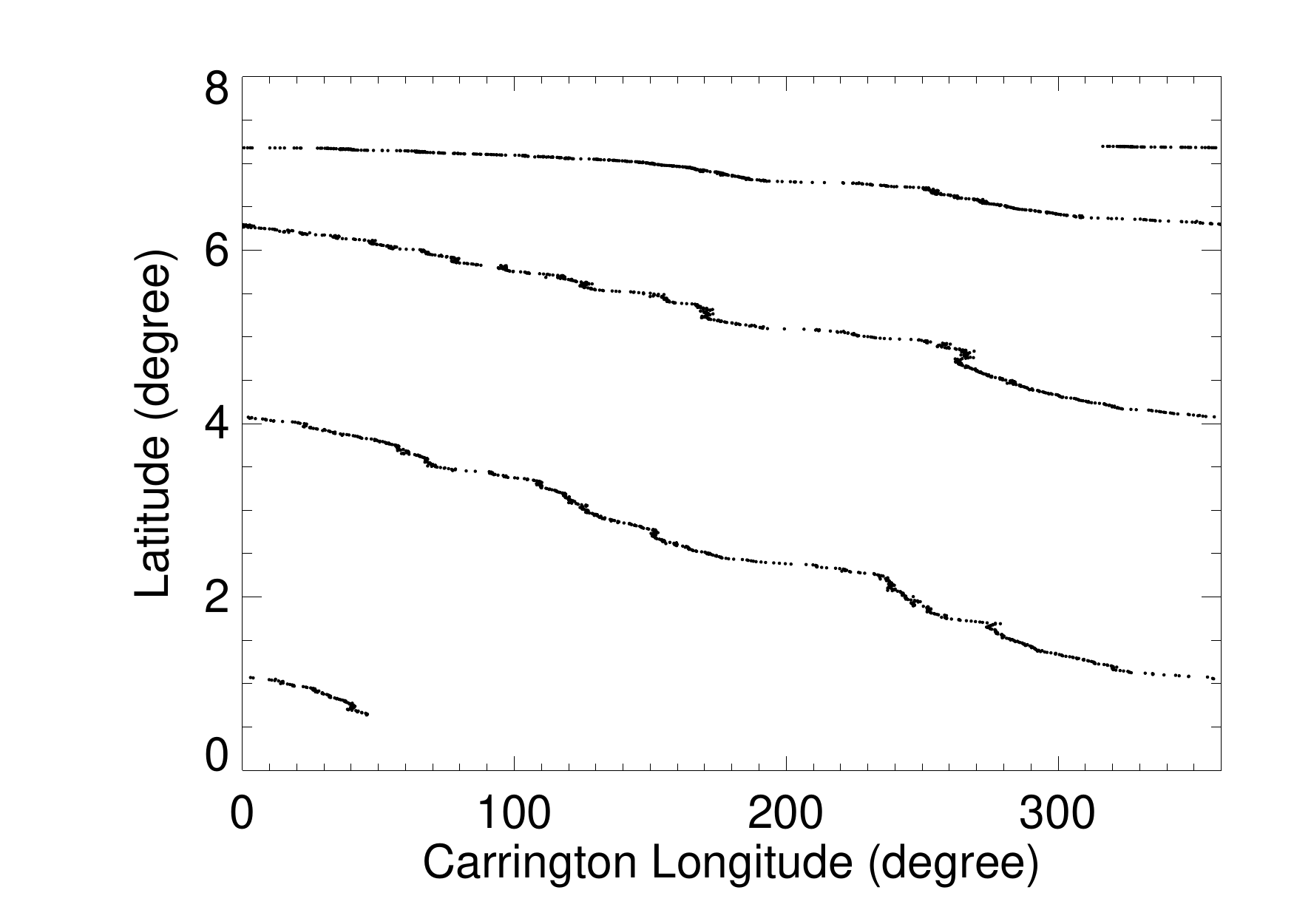}
\includegraphics[width=0.49\textwidth,clip=]{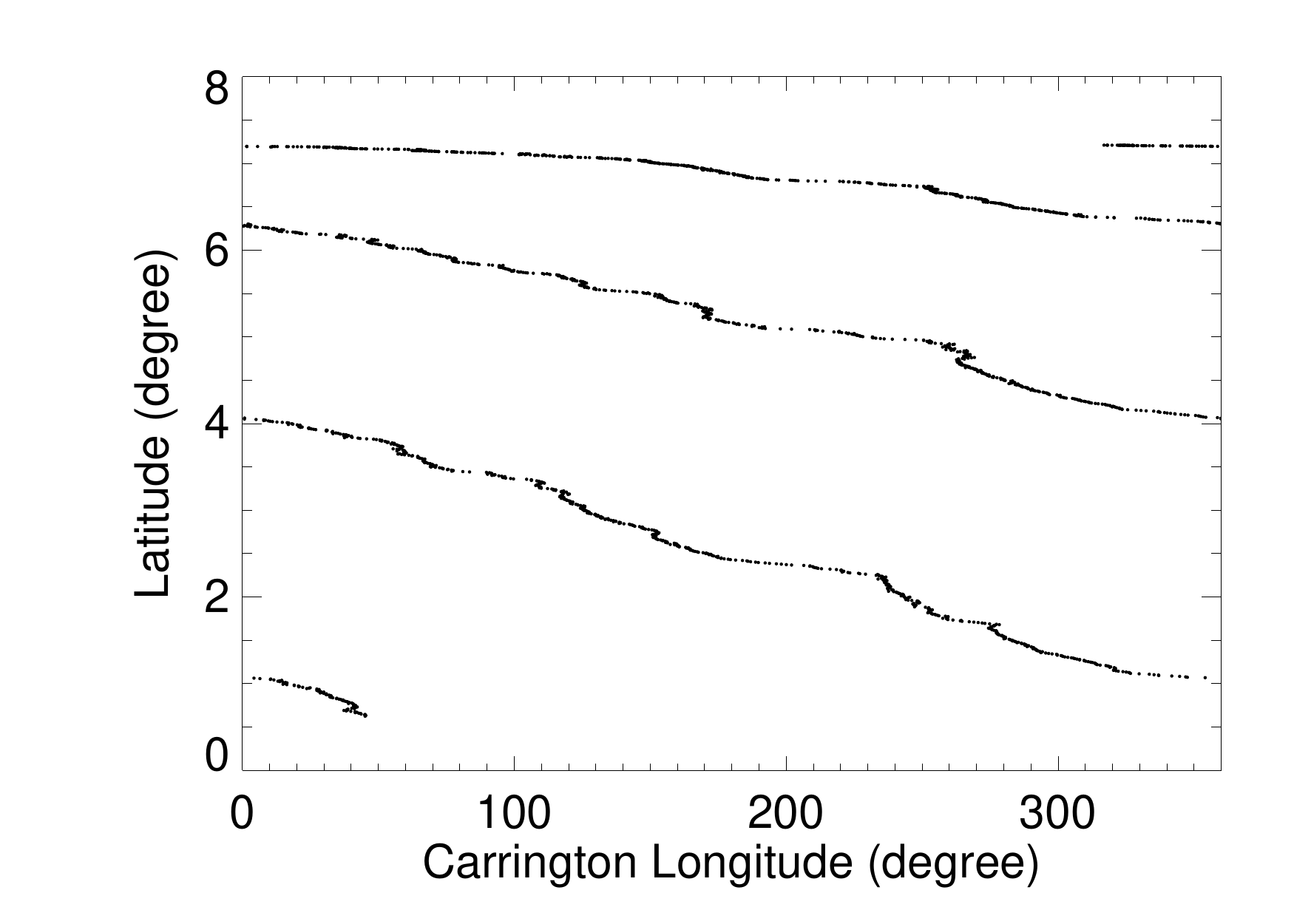}
\caption{Heliographic latitude and longitude of the source region of the solar wind
observed by the {\it Wind} (left) and the ACE (right) spacecraft during CR 2071\,--\,2073.
The solar-wind speed is assumed to be constant between the Sun and the spacecraft.}
\label{fig:source}
\end{figure}

Figure~\ref{fig:source} shows a map of the latitudes and
Carrington longitudes of the mapped-back locations on the solar
source surface of the solar wind observed by {\it Wind} and
ACE with a propagation time calculated from the observed
speeds assuming constant-speed radial flow between the source
regions and the spacecraft. The coronal-hole data are only available after September 2006, and considering that the
period (2004\,--\,2008) is in the declining phase of Solar Cycle 23,
we selected the data from CR 2071 to CR 2073 (from 11 June 2008 to 5 September 2008 at {\it in-situ}) to compare with the coronal-hole plot. We used one-hour solar-wind speed data and obtain 24
footpoints at the source surface every day. From
Figure~\ref{fig:source}, we can see that the footpoints at the
source surface jump smoothly, which agrees with the result of
\inlinecite{neu98}. Note that the two
spacecraft gradually drift to north in the studied period
\cite{mci10,neu98}.

\begin{figure}[ht]
\centering
\includegraphics[width=0.6\textwidth, clip=]{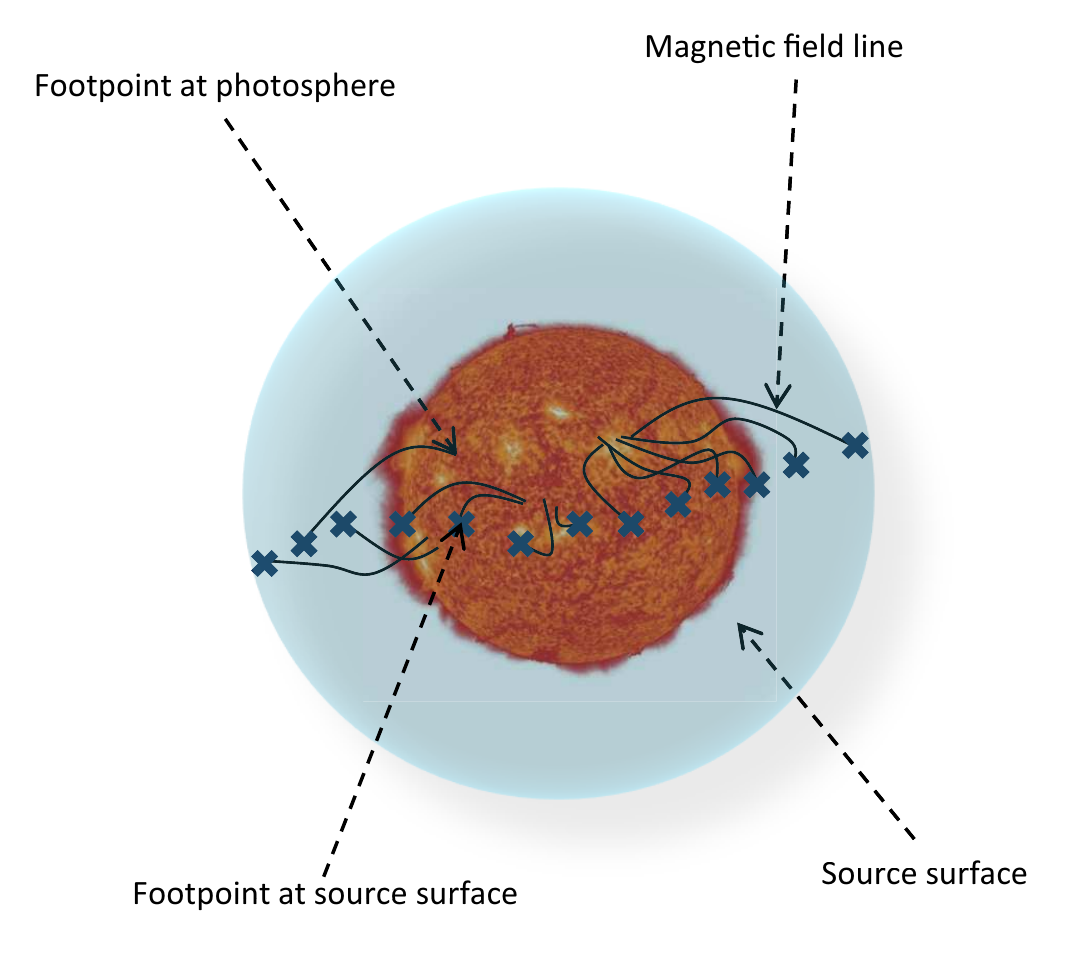}
\caption{The mapping from the source surface to the photosphere.}
\label{fig:sketchmap}
\end{figure}

\subsection{Tracking the Solar Wind Back from the Source Surface to the Photosphere}\label{daily}

The second step is to trace back the solar wind from the source
surface to the photosphere. We used the PFSS model (\inlinecite{sch69} and \inlinecite{alt69})
in this step. In this model, the coronal magnetic field is assumed
to be quasi-stationary and is described by a potential field that
can be approximated by an expansion to series of spherical harmonics.
We used the standard PFSS package included in SolarSoft
(\opencite{schrijver2001}, \citeyear{schrijver2003}). Figure~\ref{fig:sketchmap} is
a map of field lines that trace back from the source surface
(located at $2.5$ R$_\odot$) to the solar photosphere. The thick
crosses on the translucent blue sphere are the footpoints on the
source surface. The solid lines are magnetic-field lines
extrapolated using the PFSS model.

\begin{figure}[ht]
\centering
\includegraphics[width=0.95\textwidth,clip=]{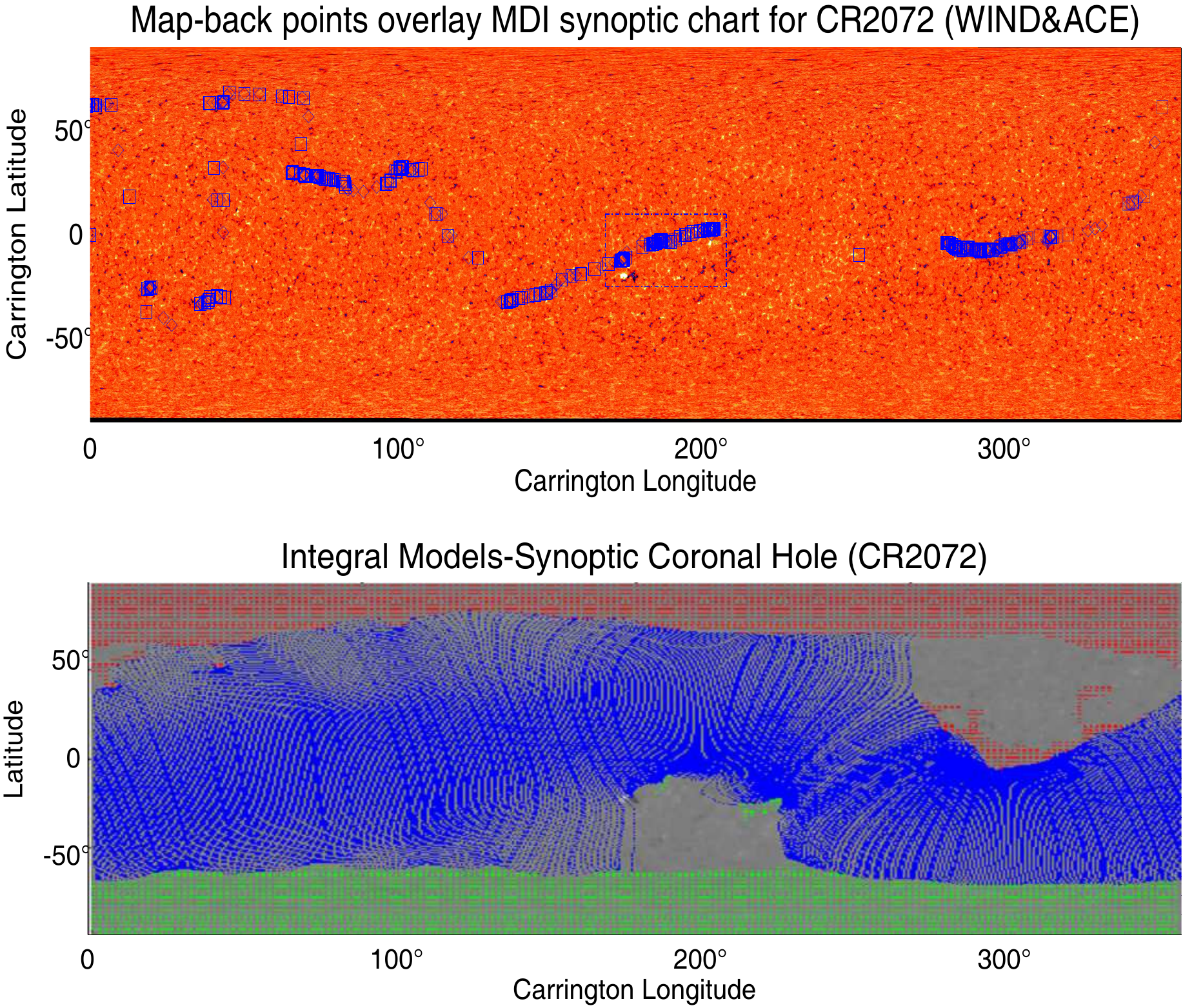}
\caption{Upper panel: Synoptic chart from SOHO/MDI during
CR 2072. The diamonds and squares show the footpoints that trace back
from the positions of source surface with {\it Wind} and
ACE data, respectively; the white dashed line represents the solar equator. The detail of the blue frame region is shown in Figure~\ref{fig:networkfp}. Lower panel: The Integral
Models Synoptic Coronal Hole from NSO/GONG, the polarities of both
the open ecliptic-plane flux and the coronal holes are indicated by
the same color code: green for positive polarity and red for
negative polarity. The tallest closed-flux trajectories are plotted in blue.}
\label{fig:footpoint}
\end{figure}

For the extrapolation, we used the SOHO/MDI synoptic chart. To
relate a footpoint to a particular magnetic cell, we superposed the
footpoints on the synoptic chart. The Carrington synoptic chart is
a collection of the center of full-disk magnetograms during one
Carrington rotation. These synoptic charts have been
reconstructed using re-calibrated magnetogram data
\cite{scherrer1995}. Figure~\ref{fig:footpoint} shows the synoptic
chart from SOHO/MDI during CR 2072. The diamonds and
squares show the footpoints that are traced back from the source
surface using the {\it Wind} and ACE data,
respectively. The footpoints do not move
smoothly across the synoptic chart. Instead, they undergo sudden
jumps on the synoptic chart, part of them overlapping on the
boundaries of the coronal holes. A large portion of the footpoints
originate from coronal holes. We compared this with the Integral
Models Synoptic Coronal Hole Plot, which is shown at the bottom of
Figure~\ref{fig:footpoint}, and found that most of the footpoints are
located either at the boundaries or in the coronal holes. This result
is consistent with those of \inlinecite{mci11} and \inlinecite{neu98}. Note
that most of the footpoints deduced from {\it Wind} are the
same as those deduced from ACE. This is expected because
the separation between ACE and {\it Wind} is small.

In the following, we use the data observed by the {\it Wind} spacecraft to
calculate the jump times of the footpoints. The jump time t, in Carrington coordinates, is equivalent to the longitudinal difference between the two adjacent footpoints divided by the solar rotation speed:
\begin{equation}
t=\frac{\left|s2-s1\right|}{V_{\mathrm{rs}}},
\end{equation}
where $s1$ and $s2$ are the longitude of adjacent footpoints, and $V_{\mathrm{rs}}$ is the rotation speed of the Sun.

Note that toward the two sides of the magnetogram, the tracing process
needs to be calculated carefully because the source region and field
line may only last a fraction of a Carrington rotation. In the two
sides of the daily synoptic chart, the magnetic field of the
region changes faster than the normal level, which may lead to variations of the
extrapolated field lines. Thus, when we track back along
these field lines, the backtracked footpoints are not static.
To extend this model, we used the daily synoptic chart as
the boundary of PFSS model. In this model, the field lines
extrapolated from the two sides of the daily synoptic chart could be
in their actual condition, therefore we can trace the field lines more
accurately. Applying this extended model, we mapped the solar
wind back to the solar surface with {\it Wind} and ACE
data, from the source surface to the solar photosphere. We found that
most of the footpoints that trace the field lines extrapolated from
the daily synoptic chart are clustered. Therefore we selected the
relatively stable daily footpoints, whose variance is less than $1
\sigma$ during one CR, to calculate the jump times of adjacent
footpoints.

\begin{figure}[ht]
\centering
\includegraphics[width=0.8\textwidth,clip=]{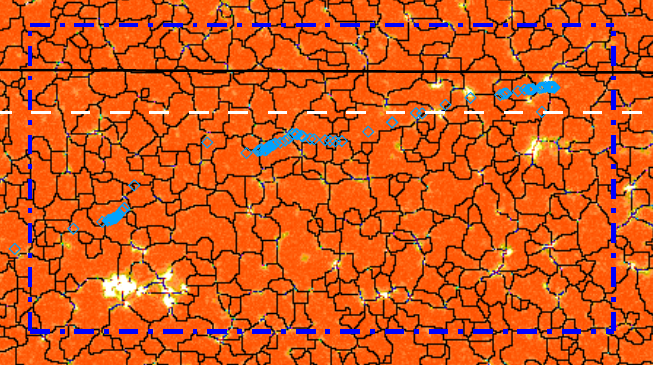}
\caption{Enlarged images for the window framed in Figure~\ref{fig:footpoint}. The longitude of the region is from
$160^\circ$ to $210^\circ$, and the latitude is from $-25^\circ$
to $10^\circ$. The white dashed line is the solar equator, the sky-blue diamonds are footpoints, that are traced back by employing
{\it Wind} data, and the black solid line is the projection
point of the {\it Wind} spacecraft during CR 2072. } \label{fig:networkfp}
\end{figure}

Figure~\ref{fig:networkfp} presents the traced-back footpoints
superimposed on the magnetic cells, which were identified with the
watershed algorithm. Some of the sky-blue footpoints are inside
the same magnetic cell; the others are at the boundary of the
cell. The jump times of all adjacent footpoints were calculated,
and we tallied the statistical results in two ways: the first way was
to examine the jump times between all adjacent footpoints,
regardless of whether they were within the same cell or in different cells; the
second way was to count only the jump times between different
cells.

\section{Results}

In the scenario where the solar wind contains numerous flux tubes
({\it e.g.} \cite{bru01,bor08,li08a,li08b}), the solar-wind plasma resides in
different flux tubes, which originate from the top of the magnetic
carpet. These flux tubes wander around their footpoints
while they propagate out during their merging or splitting. The
time scale of the merging and splitting is currently unclear, however. If the merging and splitting time scales of these flux
tubes are much longer than the propagating time of the solar wind
from the Sun to Earth, {\it i.e.} if they can survive intact beyond
100 hours, then they can be regarded as nonevolving ¡°fossil
structures¡± of magnetic cells near the photosphere. In this
case, one expects the observation of these flux tubes at 1 AU to
resemble their footpoints on the solar surface. In particular, the
crossings of flux tubes at 1 AU are expected to correspond to the jump
between magnetic cells on the solar surface. Statistically, we therefore expect that the waiting times of current sheets
are similar to the footpoint jump times between magnetic
cells at the photosphere.

\inlinecite{mia11} recently used {\it Ulysses} data to examine the
PDF of the waiting times of current sheets in the solar wind. We examine the PDF of the jump times of the
traced-back solar-wind footpoints. We also note that several
authors have reported that the lifetime of the supergranules (candidates for the magnetic cells) is about one day
\cite{wang1989,hirzberger2008}. This justifies our choice of using
one-hour data for our statistical analysis.

For the reasons discussed in Section~\ref{daily}, and because the daily
synoptic chart can reflect the transient magnetogram more
accurately, we used the daily synoptic charts to extrapolate the
magnetic-field lines and then traced back the footpoints that are
located on the source surface to the photosphere. Compared with
one Carrington synoptic chart, it is more accurate for the tracing-back process to apply about 27 daily synoptic charts in one
Carrington rotation. Because the daily tracing-back processes are
relatively independent, we calculated the jump time of adjacent
footpoints in each tracing-back process during one Carrington
rotation. Thus, this data-processing produces 26 more jump
cases than the actual physical scenario. In this statistical
analysis, we scaled the data to meet the actual scenarios by using statistical averaging.

\begin{figure}[ht]
\centering
\includegraphics[width=1\textwidth,clip=]{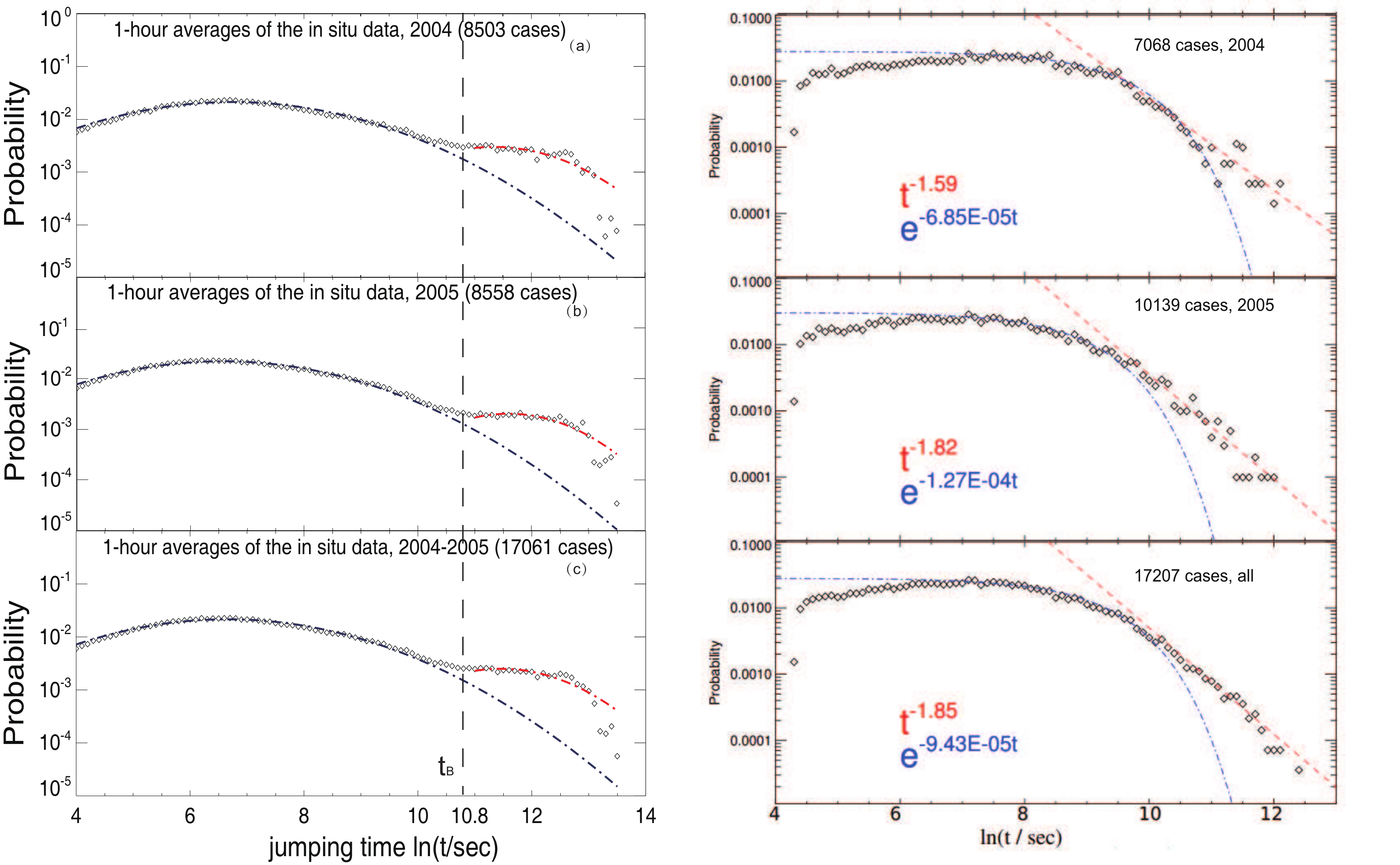}
\caption{Left panel: Statistical analysis of jump times in
different years. Panel a and b are the
jump-time analysis in 2004 and 2005, respectively. Panel
c is the jump-time analysis for all cases. Right panel:
Statistical analysis of waiting times of current sheets with all
deflection angles in different years (Miao, Peng, and Li, 2011).
From top to bottom, the three panels are the waiting-time analysis
in 2004, 2005, and all cases, respectively. The vertical dashed
line is the time at break point,$t_\mathrm{B}$. The $y$-axis is the logarithm
of the probability density, and the $x$-axis is the logarithm of the
jump time, [ln(t)], where $t$ is expressed in seconds.}
\label{fig:passtime}
\end{figure}

The left panel of Figure~\ref{fig:passtime} shows the statistical
distribution of the footpoint jump times of solar wind during
the years 2004 (a), 2005 (b), and 2004\,--\,2005 (c), respectively.
Here, the $x$-axis is the logarithm of time, and the $y$-axis is the logarithm
of the probability density. The distributions are best-fitted by two log-normal functions
(shown as blue and red curves): one at
short jump times, the other one at long jump times. The right panels of Figure~\ref{fig:passtime} show the
probability distributions of the waiting times of current sheets
with all deflection angles observed {\it in-situ} by
\inlinecite{mia11}.

In the left panel of Figure~\ref{fig:passtime}, these
three distributions are approximately log-normal distributions at
short jump times. Log-normal distributions were also found by
\inlinecite{bur99} and \inlinecite{bur01}, who obtained the
distributions of fluctuations of basic plasma fields at 1 AU
(with hour averages of the density, temperature, and speed), which
also tend to be approximate log-normal distributions for a one-year
interval. We also found that the width of the log-normal
distributions for 2004 and 2005 and the sum are almost the same. Comparing this with the right panel of
Figure~\ref{fig:passtime}, we find that on a short time-scale, the
distribution of the jump times between magnetic cells is
similar to the distribution of the waiting times of {\it in-situ}
current sheets; on a long time-scale, however, they appear to be
different. The longer waiting times of the {\it in-situ} current sheets
resemble power laws, while the footpoint jump times are
still best fitted by a log-normal.

\begin{figure}[ht]
\centering
\includegraphics[width=1\textwidth]{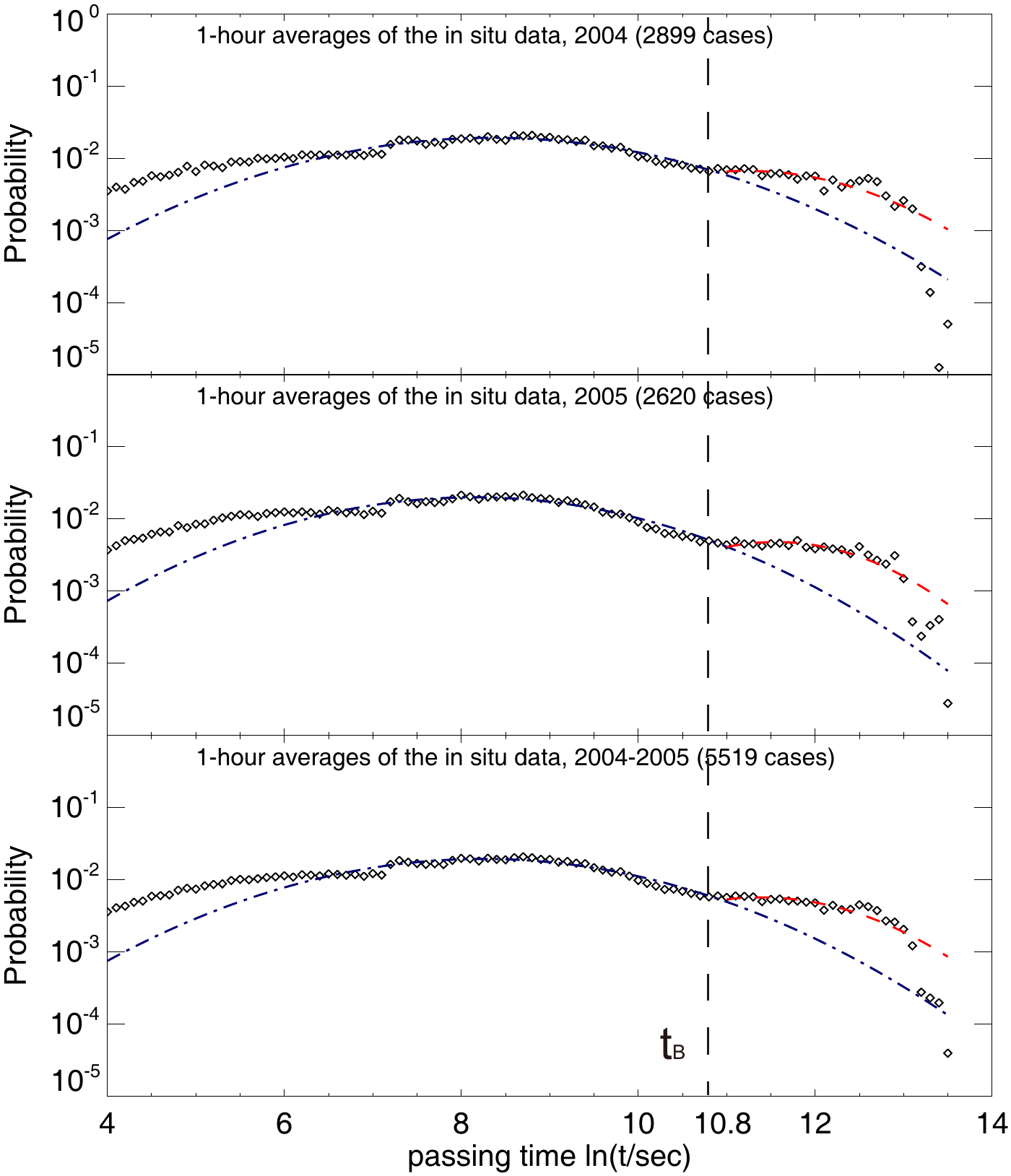}
\caption{Statistical analysis of the jump times when jumps
are crossing the boundary of the magnetic network in different years.
From top to bottom, they are for 2004, 2005, and both
years, respectively. The $y$-axis is the logarithm of the
probability density and the $x$-axis is the logarithm of jump time, [ln(t)], where $t$ is expressed in seconds.}
\label{fig:acrosspasstime}
\end{figure}

Next we examined the jump-time statistics of
footpoints that cross the boundaries of magnetic cells. These
jump times are a subset of the above study.
Figure~\ref{fig:acrosspasstime} shows their distribution.
Compared with Figure~\ref{fig:passtime}, the distributions are
similar. This suggests that there is no obvious
difference when the jump crosses the boundary or when it does not;
therefore, this result supports the scenario that the current
sheets may originate from the same mechanism. The average jump
time of footpoints that cross the boundaries of magnetic cells
($\approx e^{8.08}$ seconds) is longer than the average jump time of all
footpoints ($\approx e^{6.65}$ seconds). We note that the average jump
times crossing magnetic cells in Figure~\ref{fig:acrosspasstime}
(from top to bottom) on a long time-scale are $\approx e^{11.23}$ seconds, $\approx
e^{11.56}$ seconds, $\approx e^{11.38}$ seconds; these are roughly the same as the
average time of all footpoints in Figure~\ref{fig:passtime}. Furthermore, the curves are not fitted well on a short time-scale(when $t <
\approx e^{6.25}$ seconds) as in Figure~\ref{fig:passtime}. Indeed, the
observations showed a higher probability than the fitted curve. One
possible reason is that the footpoints are more clustered near
cell boundaries, which leads to more small-scale jump times.

\inlinecite{mia11} used
a statistical analysis of the waiting time of current sheets from
the \textit{Ulysess} data, and suggested that for all angle analyses
(\opencite{mia11}, Figure 9), the distributions behave like
exponential decays on short time-scales and power laws when $t > \lambda$
[$\lambda$ is the breakpoint of the waiting time].

In our case, we found that the result agrees with the result of waiting times of current sheets
\cite{gre09,mia11}; the PDF of the jump times of
photospheric footpoints behaves like a combination of two approximate log-normal
distributions, one on a short time-scale (when $t < t_\mathrm{B}$, where $t_\mathrm{B}$ is the break point of
the jump time) and the other on a long time-scale (when $t >
t_\mathrm{B}$). This represents that the large-angle current sheets may originate from a
different mechanism from the small-angle population. The first approximate log-normal distribution is similar
to the result of \inlinecite{mia11}; this result shows that
there are perhaps small-angle current sheets that developed
in the solar-wind MHD turbulence, as shown by \inlinecite{zhou04}
and \inlinecite{chang04}. Therefore, these current sheets may
represent the intrinsic intermittency of the solar-wind MHD
turbulence. We suggest that the second log-normal distribution is that of the current sheets with a large deflection angle; this supports the conjecture
that solar-wind turbulent fluctuations are at least in part related
to the presence of large structures of highly conducting plasma,
{\it i.e.} the flux tubes in the solar wind.

\section{Conclusion}
Previous studies by \inlinecite{bru04} and \inlinecite{mia11} have suggested that there
are two populations of current sheets in the solar wind, one of which with large deflection angles and
maybe related to flux-tube boundaries.

To examine the origin of these large-angle current sheets, we traced back the solar wind
to the solar surface and together with the magnetic-cell map using the procedure presented by
\inlinecite{hua12}, we examined the jump times between adjacent footpoints.
We identified a total of 17\,061 jumps for 2004 and 2005.
Of these, 5519 have boundary crossings.
The PDF of the jump times are shown in Figure~\ref{fig:passtime} and
Figure~\ref{fig:acrosspasstime}. These results showed that there are two populations of the jump times,
one on a short jump time-scale and one on a long jump time-scale. Both of them
can be fitted by log-normals. We also found that the average jump time
of footpoints that cross the boundaries of the magnetic network is
longer than the average jump time of all the footpoints; this is consistent
with the findings of \inlinecite{mia11}, who reported that the waiting times of current
sheets with all deflection angles are shorter than current sheets
with large deflection angles. These results support the view
that the large-angle population of current sheets may originate from
a mechanism different from that of the small-angle population, and
confirm that there might be a physical connection between the flux tube
at the solar surface and the large current sheet observed from the
{\it in-situ} data.

\begin{acks}
SOHO is a project of international cooperation between ESA and NASA.
The solar-wind data used in this work are from the {\it Wind}
and ACE teams. This work is supported by NSFC grant Nos.
11221063, 11211120147, 11103039, 11103044 and 11273030, MOST grant
No. 2011CB811401, Marie Curie Actions IRSES-295272-RADIOSUN and
the National Major Scientific Equipment R\&D Project ZDYZ2009-3.
G L's work at UA Huntsville was supported by NSF grants
ATM-0847719 and AGS-0962658. We also thank the referee for very
helpful comments.
\end{acks}

\bibliographystyle{spr-mp-sola}

\tracingmacros=2
\bibliography{sola_bibliography_example}

\IfFileExists{\jobname.bbl}{} {\typeout{}
\typeout{****************************************************}
\typeout{****************************************************}
\typeout{** Please run "bibtex \jobname" to obtain} \typeout{**
the bibliography and then re-run LaTeX} \typeout{** twice to fix
the references !}
\typeout{****************************************************}
\typeout{****************************************************}
\typeout{}}

\end{article}

\end{document}